# On the variations of acoustic absorption peak with flow velocity in Micro-Perforated Panels at high level of excitation

**Rostand Tayong, Thomas Dupont, and Philippe Leclaire**

Laboratoire de Recherche en Mécanique et Acoustique (LRMA), ISAT
Université de Bourgogne, Nevers France

The acoustic behavior of micro-perforated panels (MPP) is studied theoretically and experimentally at high level of pressure excitation. A model based on Forchheimer's regime of flow velocity in the perforations is proposed. This model is valid at relatively high Reynolds numbers and low Mach numbers. The experimental method consists in measuring the acoustical pressure at three different positions in an impedance tube, the two measurement positions usually considered in an impedance tube and one measurement in the vicinity of the rear surface of the MPP. The impedance tube is equipped with a pressure driver instead of the usual loudspeaker and capable of delivering a high sound pressure level up to 160 dB. Several MPP specimens made out of steel and polypropylene were tested. Measurements using random noise or sinusoidal excitation in a frequency range between 200 and 1600 Hz were carried out on MPPs backed by air cavities. It was observed that the maximum of absorption can be a positive or a negative function of the flow velocity in the perforations. This suggests the existence of a maximum of absorption as a function of flow velocity. This behavior was predicted by the model and confirmed experimentally.

## Introduction

Micro-perforated panels (referred to as MPPs), when associated with air cavities, are of great interest in noise reduction applications such as Helmholtz resonators. They are robust and easy to manufacture and they can be used in hostile temperature and pressure environments. Several models were developed in the linear regime to describe their surface impedance, their absorption coefficient and their transmission loss factor[1–3] . A simplified schematic of the main linear regime phenomena is depicted in Fig.1.

In the case of high sound pressure excitation (Fig. 2), it is thought that the jet formation (vorticity) at the opening modifies significantly the absorption mechanisms. By measuring the velocity in the aperture with the help of a hot wire, Ingard and Ising[4] showed that at high sound pressure levels the flow separates at the outlet orifice, forming a high velocity jet. During the inflow half-cycle, the incident flow at the inlet of the orifice is essentially irrotational but highly rotational (in form of jetting) after exiting from the outlet orifice. The acoustic particle velocity is increased sharply as the wave is squeezed into the minute perforations. The nonlinear regime implies that the acoustic properties (mainly the impedance) are dependent upon the acoustic particle velocity either in front of the panel or into the aperture. During the other half of the cycle, the flow pattern is reversed. Recently, Kraft et al.[5,6] proposed a model for MPP for high sound pressure. Their model was derived from the measurements of single degree-of-freedom liners over a wide range of sound pressure levels. It gives a prediction for the combined linear and nonlinear resistance and reactance for a face sheet. The nonlinear parameter which is considered is the discharge coefficient defined as the ratio of the actual discharge (discharge that occurs and which is affected by friction as the jet passes through the orifice) divided by the ideal discharge (without friction). The discharge coefficient value varies between 0.5 and 0.9. Kraft et al.[5,6] used a fixed discharge coefficient of 0.76 in their model assuming constant nonlinear behavior regardless of the flow conditions. Moreover, their resistive and reactive part of the plate impedance are independent of frequency essentially assuming constant orifice mass. Melling[7] derived an expression for the resistive impedance in the nonlinear regime. The discharge coefficient used by Melling is valid only for the case where the panel thickness is smaller than the orifice diameter. However, this discharge coefficient is highly frequency dependent. Later on, Maa[8] showed that the acoustic non-linearity of apertures is an external phenomenon i.e. the internal impedance is independent of the sound intensity. He suggested a nonlinear impedance term expression for MPP with small open area ratio. This expression is the ratio of the acoustic velocity inside the aperture divided by the product of sound speed and open area ratio. More recently, Hersh et al.[3] derived a model extended to multiple orifices via the open area ratio assuming continuity of volume velocity and no interactions between the holes. The discharge coefficient and some other parameters of the model were determined empirically.

The aim of this work is to propose a model for micro-perforated panels backed by an air cavity involving parameters that are easier to estimate than the discharge coefficient. This model is based on Maa's previous work on low sound pressure excitation, the use of dimensional analysis and Forchheimer's law. The first section deals with the theoretical analysis of Maa's MPP model in the linear regime. The next section presents the derivation of the proposed model. In this section, a dimensionless parameter involved in the MPP behavior is introduced. An experimental and theoretical analysis shows that the maximum of absorption can be a positive or a negative function of the flow velocity in the perforations, suggesting the existence of a maximum of absorption as a function of flow velocity. This behavior was predicted by the proposed model and confirmed experimentally. The final section describes the experimental setup, and offers comments on the results.

# I. Linear regime model

The main mechanism of absorption in the linear regime is the conversion of the acoustical energy into heat. In this regime (low sound pressure and velocity amplitudes), if the dimensions of the MPP (diameter of holes, holes separation, thickness) are small with respect to the impinging acoustic wavelength, and if the aperture dimension (diameter of holes) is of the order of the viscous and thermal boundary layers thicknesses (Fig. 1), the major part of the acoustical energy is dissipated through viscous and thermal effects.

## Maa's linear model

Based on the theory and equations of acoustical propagation in short and narrow circular tubes, Maa[1] derived an equation of aerial motion given for one perforation by

$$j\omega\rho u - \frac{\eta}{r}\frac{\partial}{\partial r}\left[r\frac{\partial}{\partial r}u\right] = \frac{\Delta p}{h}, \qquad (1)$$

where $p$ is the pressure drop across the tube, h the length of the tube (which corresponds to the thickness of the MPP), $\eta$ the dynamic viscosity, $\rho_0$ the density of air, $\omega$ the angular frequency, $r$ the radial coordinate and $u$ the particle velocity in the perforation.

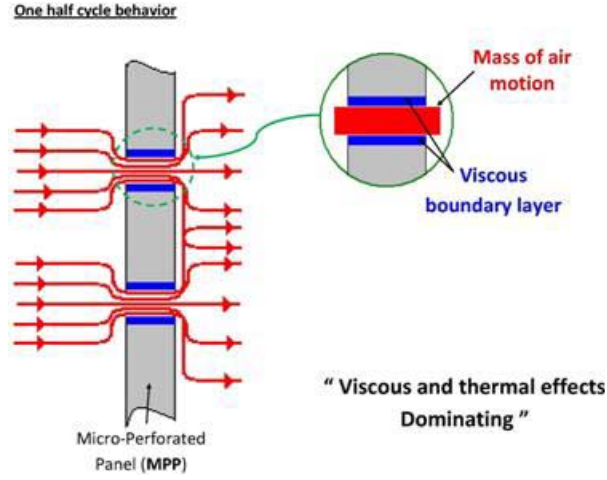

FIG. 1. Simplified schematic of the linear regime (low sound pressure levels).

By solving the equation with respect to the velocity, the specific acoustic impedance of the short tube defined as the ratio of p to the average velocity < u > over a cross-sectional area of the tube is given by :

$$Z_{perf} = \frac{\Delta p}{<u>} = j\omega\rho_0 h \left[1 - \frac{2}{x\sqrt{-j}} \frac{J_1(x\sqrt{-j})}{J_0(x\sqrt{-j})}\right]^{-1}, \qquad (2)$$

where $x = d.\sqrt{\frac{\omega\rho_0}{4\eta}}$ is a constant defined as the ratio of orifice diameter d to the viscous boundary layer thickness of the air in the orifice, $J_0$ and $J_1$ the Bessel functions of the first kind of orders 0 and 1, respectively. An approximation of the above equation on the Bessel functions and valid for narrow tubes was given by Maa1 as :

$$Z_{perf} = \frac{32\eta h}{d^2}\sqrt{1+\frac{x^2}{32}} + j\omega\rho_0 h\left[1+\frac{1}{\sqrt{3^2+x^2/2}}\right]. \qquad (3)$$

Due to the end radiating effects at the aperture, an end correction factor proposed by Rayleigh should be taken into account twice (once for each end). This correction is important when the perforation diameters are greater or of the order of the plate thickness[4]. The radiating impedance for the end correction is given by :

$$Z_{ray} = \frac{(k_0 d)^2}{8} + j\frac{8k_0 d}{6\pi}, \qquad (4)$$

where $k_o$ is the wave number. The effect of the vibration of the air particles on the baffle in the vicinity of the aperture increases the thermo-viscous frictions. To take this effect into account, Ingard and Labate[9] proposed an additional factor on the resistive part of the tube impedance. This resistance is given by :

$$R_S = \frac{1}{2}\sqrt{2\omega\rho_0\eta}. \qquad (5)$$

Under a certain number of assumptions, a process of homogenization can be applied, providing an expression of the impedance for multiple perforations. The minimal distance between perforations must be greater than their diameters and smaller than the wavelength and so it is possible to consider that there is no interactions between the apertures and the absorption mechanism is dominant. The MPP must be thinner than the wavelength to insure the continuity of the velocities on both sides of the plate. Within these assumptions, the visco-thermal interactions between the fluid the solid are taken into account through a viscosity correction function[1]

The total impedance of the MPP is then given by the impedance of one perforation divided by the open area ratio φ :

$$Z_{MPP} = \frac{Z_{perf} + 2Z_{ray} + 4R_S}{\Phi} \qquad (6)$$

## II. Model for the impedance of MPP at high sound pressure levels

### A. Variation of the impedance with MPP geometrical parameters

From experimental measurement using hot wires, Ingard and Ising[4] have observed that over a half period of the propagation of the high amplitude sound wave, the incident flow is irrotational while the outflow is a highly rotational jetting (Fig. 2). During the other half period the flow pattern is reversed. According to Ingard[10], air current losses energy at the inlet due to friction on the panel surface along which part of the air current has to move when it is squeezed into the small area of the tube. They have also observed that the orifice resistance varies linearly with particle velocity.

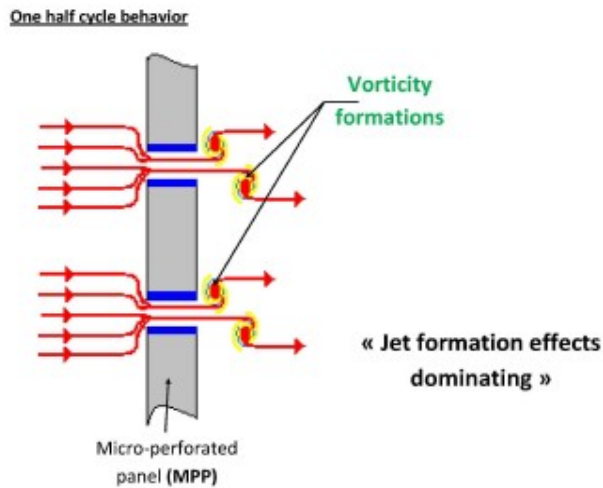

FIG. 2. Simplified schematic of the nonlinear regime (High sound pressure levels).

From the semi-empirical approach of Ingard[10], we investigated a model based on dimensional analysis[11] in order to study the influence of the MPP geometrical parameters. This model is found to be consistent with including the Forchheimer nonlinear flow regime in the linear model. This approach was used by Auregan and Pachebat[12] for the study of nonlinear acoustical behavior of rigid frame porous materials.

Strictly speaking, the characteristic impedance is not defined for nonlinear wave propagation. However, if the harmonic distortion is not too high (first harmonic at much higher amplitude than the following harmonics), it is possible to define an impedance. Ingard and Ising[4] proposed an expression where the real part of the impedance is given by

$$Re\{Z_{MPP}\}_{NL} = Au + B \ , \qquad (7)$$

where Re {$Z_{MPP}$} is the MPP resistance, $u$ the velocity in the perforation, A and B are constants. It is found that the variation of the reactance with $u$ is fairly moderate. Melling[7] explained that this remark about the nonlinear reactance (imaginary part of the impedance) is considered valid for thin plates. He stated that the reactance tends to an asymptotic limiting value (this would be confirmed by our experimental results) of approximately one-half the linear regime value.

Dividing $u$ by the speed of sound $c_0$, the Mach number $M = u/c_0$ is introduced and this expression can be written in a normalized form as:

$$Re\{z_{MPP}\}_{NL} = aM + b \ , \qquad (8)$$

where $z_{MPP}$ is the normalized impedance, a and b are now dimensionless parameters that will depend on the MPP geometrical features and on the fluid constants. This was already noticed by Maa[8] in a recent work in which he observed that $a$ is inversely proportional to the open area ratio $\phi$ while Hersh[3] observed that the resistance is dependent on the ratio $h/d$.

From the dimensional analysis[11], it is found that the variation seems to be a more complicated combination of these two behaviors :

$$a = K \left(\frac{d}{h}\right)^l \left(\frac{h\rho_0 c_0}{\mu}\right)^m \phi^n \ , \qquad (9)$$

where $K$, $l$, $m$ and $n$ are constants. However, we tend to believe that the constant $K$ is quite related to the shape of the aperture edge and to the material itself. The influence of the edges shapes on the resistive part of the orifice impedance of a cylindrical tube has been studied by Atig et al.[14]

B. Constant determination from the Forchheimer nonlinear flow model in the perforations

The study of rigid frame porous materials at high sound pressure levels by Auregan and Pachebat[12] makes use of Forchheimer's law[13], which states that for high Reynolds number (Re larger than unity), the flow resistivity increases linearly with the Reynolds number. Since the Reynolds number is proportional to the aperture diameter, the resistivity (also seen as the real part of the impedance per unit thickness) is directly proportional to the diameter. In fact, from expression (9), by considering $l = m = 1$ and $n = -1$, one easily finds the result given by Auregan and Pachebat[12] if the viscous characteristic length is taken equal to the perforation radius (case of cylindrical pores).

Finally, the model for the coefficient of the real part of the MPP impedance is

$$Re\{z_{MPP}\}_{NL} = aM + b \qquad (10)$$

$$a = K \left(\frac{d}{h}\right)^1 \left(\frac{h\rho_0 c_0}{\mu}\right)^1 \phi^{-1} \ , \qquad (11)$$

a can be further simplified as

$$a = K\frac{dc_0}{v\phi},\qquad(12)$$

where v is the kinematic viscosity. Now, still following Auregan and Pachebat[12] work, expression of b is related to the low sound excitation resistance value of the MPP. They observed 3 regimes : the linear, the transition and a weakly nonlinear regime. An appropriate expression is given by :

$$b = (1+\delta)Re\{z_{MPP}\}_{linear}.\qquad(13)$$

In fact, $b$ is the intercept with vertical axis (real part of impedance) and $\delta$ is a dimensionless parameter adapted to describe the high sound pressure regime. The parameters $K$ and $\delta$ are to be determined experimentally.

III. Absorption coefficient of a MPP backed by a cavity

Under the same assumption on harmonic amplitudes, it is possible to define a cavity impedance for high sound pressure levels and it is given by the usual expression (normalized form) :

$$z_c = -jcot(k_0 D_c),\qquad(14)$$

where Dc is the air gap thickness and the normalized surface impedance of the total system (MPP coupled to an air cavity) is given by the superposition of the two impedances[1,4] :

$$z_s = z_{MPP} + z_c.\qquad(15)$$

The reflection coefficient is given by the usual formula

$$R = \frac{z_s-1}{z_s+1}\qquad(16)$$

and the acoustic absorption coefficient can be calculated from

$$\alpha(\omega) = 1 - |R(\omega)|^2.\qquad(17)$$

When the real and the imaginary part of the total impedance zs are separated, another expression of the absorption coefficient can be given by :

$$\alpha = \frac{4Re\{z_s\}}{(1+Re\{z_s\})^2+(Im\{z_s\})^2}\qquad(18)$$

IV. Maximum absorption as a function of Mach number

By inserting the expression of the real part of the MPP impedance in the equation for the absorption coefficient we have

$$\alpha = \frac{4(aM+b)}{(1+aM+b)^2+(Im\{Z_s\})^2}\qquad(19)$$

The maximum of absorption is obtained for Im{zs } = 0 :

$$\alpha_M = \frac{4(aM+b)}{(1+aM+b)^2}\qquad(20)$$

This expression is now differentiated with respect to M in order to study the variations with the Mach number (and implicitly with the incident sound pressure level)

$$\frac{\partial \alpha_M}{\partial M} = \frac{4a(1-aM-b)}{(1+aM+b)^3} \qquad (21)$$

This result shows that a critical value of the Mach number exists for which the absorption coefficient is extremum (Fig. 3). Since $a = 0$, this last expression provides a limit Mach number $M_c$ given by:

$$M = \frac{1-b}{a} \qquad (22)$$

From the study of the variations of the absorption coefficient function, it is found that the peak of absorption (maximum of absorption with respect to frequency) increases with the Mach number, reaches a maximum as the Mach number approaches its critical value and then decreases for M increasing beyond the critical value $M_c$. It is worth noticing that Maa[8] predicted this behavior. In the present article, we propose a more refined model, we define a critical value for the Mach number and include experimental data.

Evidently, this behavior can be observed only if the critical value is above the linear/nonlinear regime limit. Indeed, the experimental results will show that in some cases, a value for $M_c$ will not be identified if it is located in the linear range. In this case the MPP absorption peak will only decrease with the increasing sound pressure level.

It is also worth noticing that $\alpha_M (M_c) = 1$ for any MPP with Mc located in the nonlinear domain. This result shows potential applications in the design of MPPs.

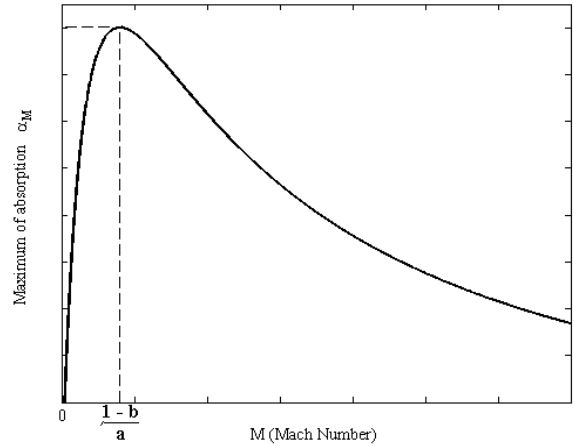

FIG. 3. Absorption coefficient at the resonance frequency versus the acoustical Mach Number in front of the MPP.

# V. Experiments

A. MPP samples

The measurements are performed on strong copolymer-made MPP and steel-made MPP (Fig. 4 and Table I). All the sample panels have an external diameter of 100 mm and all the holes are well separated from each other (no interactions between the apertures) and are evenly distributed over the panel area. The mounting conditions of the MPP inside the tube are closer to the clamped conditions than to the simply supported conditions.

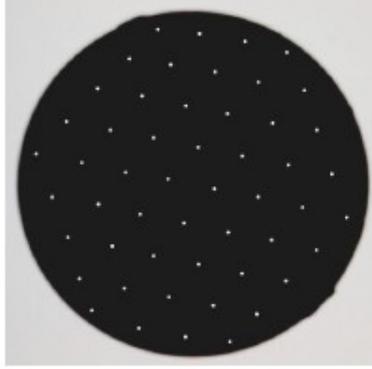

FIG. 4. Perforated panel Sample.

TABLE I. Micro-perforated panel characteristics.

|      | h (mm) | d (mm) | Φ (%) | Density (Kg/m$^3$) | Young Modulus (Pa) | Material |
|------|--------|--------|-------|--------------------|--------------------|----------|
| MPP1 | 2.2    | 1.0    | 2.2   | 900                | $0.49 \times 10^{10}$ | Strong copolymer |
| MPP2 | 2.0    | 1.0    | 0.8   | 900                | $0.49 \times 10^{10}$ | Strong copolymer |
| MPP3 | 2.0    | 0.7    | 1.94  | 7800               | $21 \times 10^{10}$   | Steel |

B. Impedance tube and data acquisition

A schematic of the impedance tube used is shown in Fig. 5. It is a rigid circular plane-wave tube with a diameter of 100 mm. Plane wave propagation is assumed below the cut-off frequency (1.7 KHz). At the left hand side, a compression driver JBL model 2450J is mounted as the source of excitation. A transition piece provides a continuity transition between the circular section of the compression driver and the circular cross section of the plane-wave tube. At the right hand side of the tube, a soundproof plunger is used as the rigid backing wall. By moving the plunger along the longitudinal axis of the tube, one is able to create an air cavity behind the MPP sample. The MPP sample is mounted between the speaker and the plunger. Three ¼ microphones are used to perform the signal detection. Two microphones are used to calculate the surface impedance of the sample by the standard impedance tube measurement technique[15]. The third microphone (reference micro in Fig. 5) acts as a reference microphone to get the level of pressure at the sample surface. When performing high sound pressure measurements inside an impedance tube, it is important to check the resulting standing wave. In fact, up to a certain level, depending on the frequency excitation and the sample parameters, the standing wave seems to saturate and therefore the linear propagation hypothesis would no longer be valid. Effects of bifurcation may occur. These phenomena of saturation and bifurcation were observed and shown by Maa and Liu[16]. A first excitation is done with a periodic random noise signal in order to have a general view of the absorption coefficient curve and locate the viscous peak position. The result is used to determine the resonant frequency and to perform a second excitation (sine excitation) at frequencies concentrated around the viscous absorption peak(s). The amplitude of the source is adjusted such that the sound pressure level measured by the reference microphone is set at the desired level. The SPL (Sound Pressure Level) is varied from 90-155 dB at the face of the MPP monitored using the reference microphone. A phase and amplitude calibration method is used to correct the transfer function between the measurement microphones.

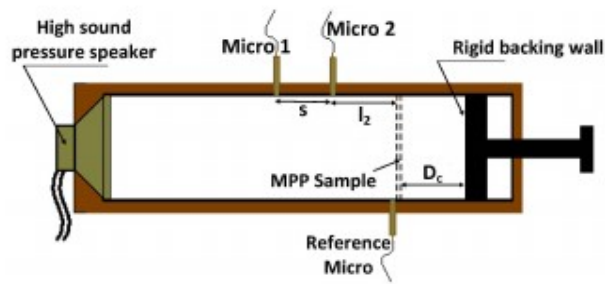

FIG. 5. Schematic of the impedance tube used for the measurements.

TABLE II. Dimensionless parameters of the micro-perforated panels.

| | K | δ |
|---|---|---|
| MPP1 | 1.43 × 10$^{-3}$ | 2.024 |
| MPP2 | 1.43 × 10$^{-3}$ | 3.868 |
| MPP3 | 1.56 × 10$^{-3}$ | 0.956 |

The measurement are carried out at high sound pressure levels. However, assuming the plane wave hypothesis, one can measure the pressures and velocities on any section of the tube using the two microphone method. See for instance Dalmont[17] for more details. The acoustic velocity on the panel surface is given by:

$$u = j\frac{p_1}{Z_0}\frac{H\cos(k_0 l_1)-\cos(k_0 l_2)}{\sin(k_0 s)} \quad (23)$$

where $l_1 = s+l_2$ as in Fig. 5. $Z_c$ is the characteristic impedance of air, $k_0$ is the wave number, $p_1$ is the pressure on microphone 1, and $l_1$ (resp. $l_2$) is the distance from microphone 1 (resp. 2) to the panel sample. The values of the velocity shown in the experimental results are the viscous peak corresponding particle velocity.

# VI. Results and comments

In this section, the measurements are performed taking a single MPP with an air cavity behind and a rigid wall. The dimensionless parameters K and δ used to fully determine expression (8) are given in Table II)

Fig. 6 shows the comparison between experimental results and the present model simulations for the resistance as a function of the Mach number for all the MPP samples. The air cavity depth is 50 mm. The experiments and the present model are in good agreement for the high excitation levels. This result points out the fact that for high sound pressure levels, the dependency of the resistance and the Mach number is linear. The slope is different from one MPP to another, revealing therefore the fact that this slope, in the same measurement conditions, depends on the MPP geometrical parameters. This result also shows that the value of constant K is related to the type of material (1.43 × 10$^{-3}$ for the copolymer samples and 1.56 × 10$^{-3}$ for the steel sample). As already mentioned, we tend to believe that the constant K is quite related to the shape of the aperture edge and to the thermal properties of the material considered.

Fig. 7 shows the comparison between experimental results and the present model simulations for the maximum absorption coefficient (viscous peak) versus the Mach number for all the MPP samples. The air ca-vity depth is 50 mm. The experiments and the simulation are in fairly good agreement. This confirms the fact that depending on the value of the limit Mach number, with the increase of sound levels, the viscous peak will in a first phase rise up to a maximum value before decreasing. This limit Mach number point is observable on MPP1 and MPP3 but not on MPP2. In fact, if this limit Mach number is low enough, the absorption peak will solely decrease with the increase of sound pressure levels.

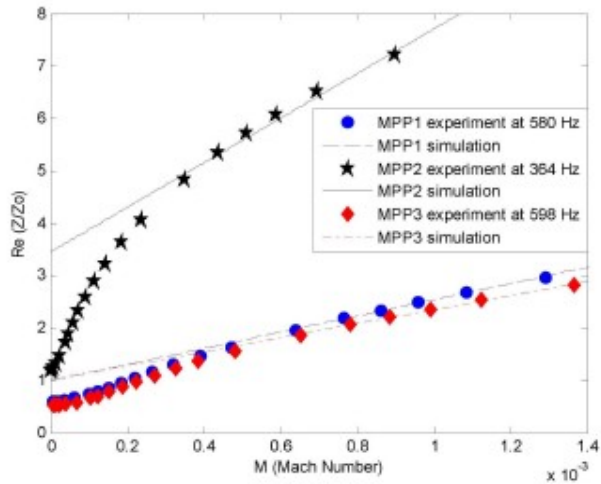

FIG. 6. Surface impedance resistive part as a function of the Mach number in front of the MPP sample. Air cavity depth of 50 mm.

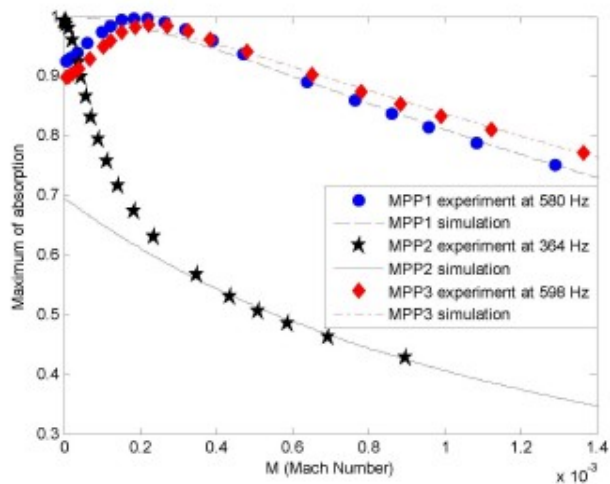

FIG. 7. Absorption coefficient at resonance as a function of the Mach number in front of the MPP sample. Air cavity depth of 50 mm.

Fig. 8 shows the comparison between experimental and simulations results (Maa[8], Hersh[3] and the present model) for the absorption coefficients of MPP1 at 145 dB in front of the panel (reference microphone) in the [200 - 1600 Hz] frequency range for an air cavity depth of 50 mm. The simulation of the present model and the measurement are in very good agreement. The first peak (around 564 Hz) is the viscous peak and the second peak (around 900 Hz) is the result of the structural response of a panel coupled to an air cavity. This structural response is described in the appendix. The presented models and the measurements are in fairly good agreement except for the Hersh high sound model which does not fit well with the experiment results.

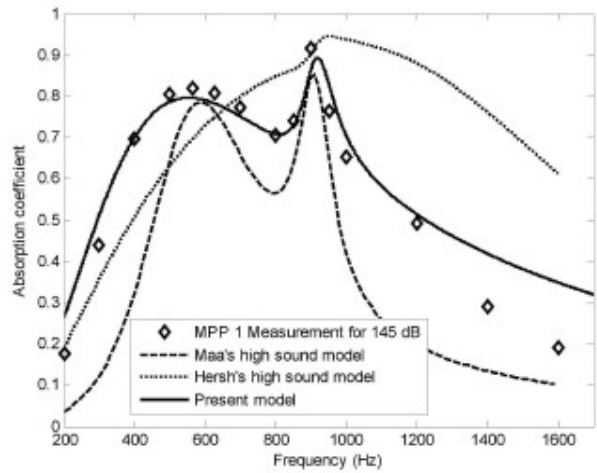

FIG. 8. Comparison between the absorption coefficients of MPP1 at 145 dB (U=0.325m/s) on the reference microphone. Air cavity depth of 50 mm.

Fig. 9a and Fig. 9b show the comparison between experimental and simulations results of the surface impedance (resistance (a) and reactance (b)) of MPP1 at 145 dB in front of the panel (reference microphone) in the [200 - 1600 Hz] frequency range for an air cavity depth of 50 mm. The present model and the measurements are in very good agreement. On Fig. 9a, Maa's model underestimates the nonlinear resistance whereas Hersh's model agrees well with the experimental results except around the structural response frequency. On Fig. 9b, the models are in good agreement with the experimental results except for Hersh's model.

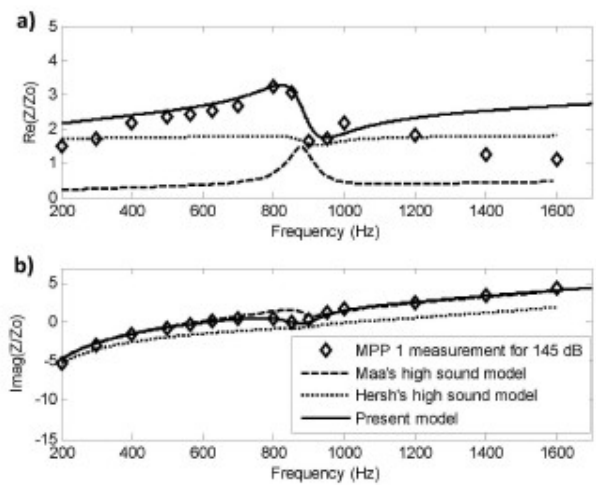

FIG. 9. Comparison between the surface impedances (resistance (a) and reactance (b)) of MPP1 at 145 dB (U=0.325m/s) on the reference microphone. Air cavity depth of 50 mm.

Fig. 10 shows the comparison between experimental and simulation results (Maa[8], Hersh[3] and the present model) for the absorption coefficients of MPP3 at 145 dB in front of the panel (reference micropho-ne) in the [200 - 1600 Hz] frequency range for an air cavity depth of 40 mm. The present model and the measurement results are in very good agreement.

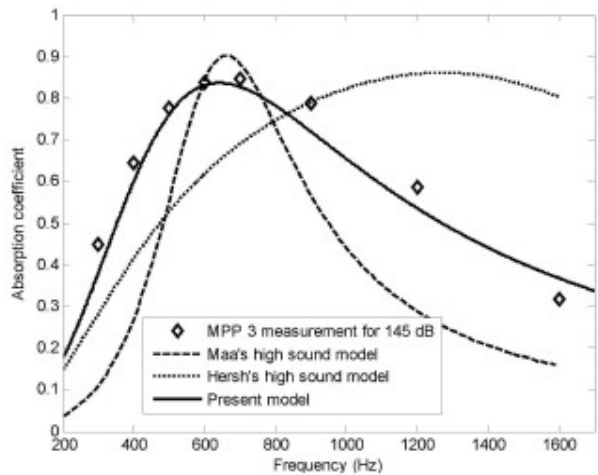

FIG. 10. Comparison between the absorption coefficients of MPP3 at 145 dB (U=0.337m/s) on the reference microphone. Air cavity depth of 40 mm.

Fig. 11a and Fig. 11b show the comparison between experimental and simulations results for the surface impedance (resistance (a) and reactance (b)) of MPP3 at 145 dB in front of the panel (reference microphone) in the [200 - 1600 Hz] frequency range for an air cavity depth of 40 mm. On Fig. 11a, for the resistance part, below 800Hz, the present model and the measurement results are in very good agreement. Yet beyond 800 Hz, the agreement is not good. This may imply considering a certain high frequency-dependency of the nonlinear parameter for a more accurate prediction. However, as mentioned, for relatively high frequency, the absorption seems to be much more influenced by the imaginary part of the impedance. Maa's model underestimates the measurement results while Hersh's model tendency is good compared to the measurement results. On Fig. 11b, for the reactance part, the present model and Maa's model both agree accurately with the measurement results.

## Conclusion

A model for the high sound pressure of Micro-Perforated Panels (MPP) when backed by an air cavity has been proposed and tested experimentally. A dimensionless parameter and a limit Mach number were found and used. These are suitable to predict the impedance of the system at high sound pressure levels. The results and the analysis showed that the model and the experiments are in good agreement. The theoretical work revealed the fact that with the increase of sound pressure level (or Mach number), the viscous absorption peak will in a first phase rise to a maximum value and then decrease in a second phase. Experimentally, it was noticed that this latter result can clearly be observed if the value of the limit Mach number is above the linear regime limit of the MPP. If the limit Mach number is below the linear regime limit, the viscous absorption peak will only decrease with increasing sound pressure level. Finally, it was shown that MPP are very sensitive to the incident sound pressure. Thermal effects around the MPP system tends to play an important role with the increase of sound intensities. Further work will consist in investigating these effects and to properly understand their influence on the MPP impedance for high sound intensities. It may also be interesting to investigate the apertures interactions when submitted to relatively high sound pressure amplitudes.

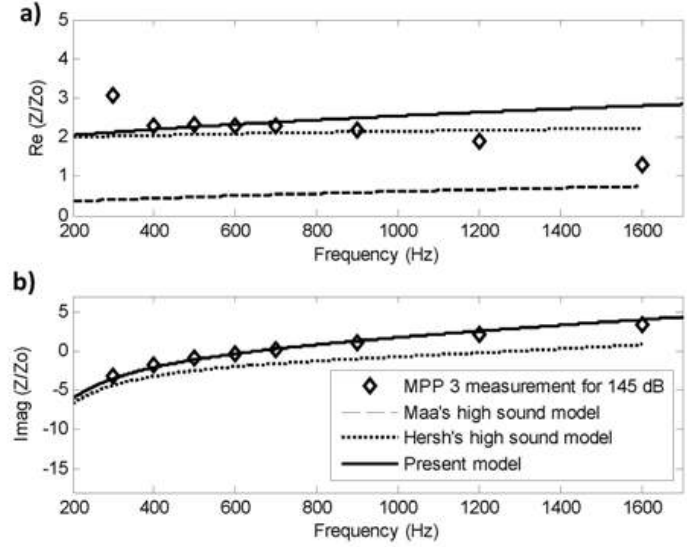

FIG. 11. Comparison between the surface impedances (resistance (a) and reactance (b)) of MPP3 at 145 dB (U=0.337m/s) on the reference microphone. Air cavity depth of 40 mm.

## Acknowledgments

The support grant for this work was provided by the Conseil Régional de Bourgogne.

## APPENDIX A: model including the structural behavior of the panel

In the absorption test of the copolymer micro-perforated panel, an additional sound absorption peak was found around the frequency of 900Hz due to the panel vibration first mode. Theoretical and experimental studies were done by Ford and McCormick[18] and also Frommhold et al.[19] on panel absorbers without perforations. From these studies, it was found that this additional absorption peak is due to the panel-cavity resonance. Ford and McCormick[18] gave an expression of the structural impedance of such system (membrane coupled to an air cavity). This expression is considered and used in this paper when the structural resonant frequency of the panel is not beyond the frequency range of study (200-1600Hz). Only the first structural mode is considered since the other structural frequency modes are always out of the frequency range studied. Since the MPP backed by an air cavity system can be considered analogous to an electrical circuit, it is well assumed that the normalized acoustic impedance of the whole system Zs is given by:

$$Z_s = \frac{Z_{MPP} Z_{vib}}{Z_{MPP} + Z_{vib}} + Z_c , \qquad (A1)$$

where $Z_{MPP}$ is the MPP surface impedance. $Z_c$ is the air cavity impedance. $Z_{vib}$ is the structural impedance of an air cavity backed membrane (without perforations) well described by Ford and McCormick[18] as:

$$Z_{vib} = \frac{D_{rig} B_{mn} \xi}{\omega h^4} + j\omega M_p A_{mn} + \frac{1}{j\omega}\left(\frac{D_{rig} B_{mn}}{h^4} + \frac{\rho_0 c_0^2}{D_c}\right) , \qquad (A2)$$

where $A_{mn}$ and $B_{mn}$ are the modal constants, Mp the mass per unit area of panel, *h* is the lateral dimension (thickness of the panel), $D_c$ is the air cavity depth and $D_{rig}$ the bending stiffness of the panel. Where

$$D_{rig} = \frac{Eh^3}{12(1-v_p^2)} \quad (A3)$$

where E is the Young modulus (Table I), ξ=10% and $v_p$ =0.3. Since our MPP mounting conditions are rather close to clamped-clamped, $A_{mn}$ =2.02 and $B_{mn}$ =2640. (See table I for the other parameters). An important remark to mention in this section is about Lee et al.[20] observations that if the forcing frequency (viscous peak) is higher than the first structural resonant (structural peak), then the structural vibration degrades the absorption. And in contrast, if the forcing frequency is lower than the first structural resonant frequency, the structural vibration enhances the absorption.